\documentclass[12pt,a4paper]{article}
\usepackage{epsfig,amsmath,amssymb,subfigure,placeins}
\usepackage[colorlinks=true,linktocpage=true,linkcolor=blue,citecolor=blue]{hyperref}
\usepackage{color}
\newcommand{\old}[1]{}

\newcommand{\be}{\begin{equation}}
\newcommand{\ee}{\end{equation}}
\newcommand{\ba}{\begin{eqnarray}}
\newcommand{\ea}{\end{eqnarray}}
\newcommand{\bi}{\begin{itemize}}
\newcommand{\ei}{\end{itemize}}

\begin{document}
\begin{flushright}
{\normalsize
}
\end{flushright}
\vskip 0.1in
\begin{center}
{\large{\bf Heavy Quark Potential at Finite Temperature in a  Dual Gravity Closer to Large N QCD}}
\end{center}
\vskip 0.1in
\begin{center}
Binoy Krishna Patra and Himanshu Khanchandani\\ 
{{\it Department of Physics, Indian Institute of Technology Roorkee, Roorkee 247 667, India} }
\end{center}
\vskip 0.01in
\addtolength{\baselineskip}{0.4\baselineskip} 


\begin{abstract}
In gauge-gravity duality, the heavy quark potential at finite temperature is usually 
calculated with the pure AdS background, which does not capture the renormalization group 
(RG) running in the gauge theory part. In addition, the potential does not contain any 
confining term in the deconfined phase. Following the  Klebanov-Strassler 
geometry, we employ a geometry, which captures the RG flow similar to QCD, to obtain the 
heavy quark potential by analytically continuing the string configurations into the 
complex plane. In addition to the attractive terms, the obtained potential has confining 
terms both at $T = 0$ and $T \neq 0 $, compared to the calculations usually done in the 
literature, where only the Coulomb-like term is present in the deconfined phase. The 
potential also develops an (negative) imaginary part above a critical separation, $r_c (= 
0.53 z_h )$. Moreover, our potential exhibits a behavior different from the usual Debye 
screening obtained from perturbation theory.
\end{abstract}


\vspace{1mm}

\section{Introduction}
The heavy quarks produced in the early stage of relativistic 
heavy-ion collisions (HIC) is one of the crucial probes to the 
medium formed at later stage of the collision, 
known as quark-gluon plasma (QGP). Matsui and Satz~\cite{Matsui:1986dk} 
first proposed the idea of Debye screening of 
the potential between a heavy quark and a heavy antiquark, which 
causes the suppression of the yields of heavy quarkonium 
states in HIC~\cite{Karsch:1987pv}. 
Since then many efforts have been devoted to understand the 
change of properties of $Q \bar {Q}$ states in the deconfined medium, using 
either the non-relativistic calculations based upon some
effective (screened) potential \cite{Dig01,Shu04,Won07,Cab06,Alb07,Moc07}
or the lattice calculations of quarkonium spectral functions
\cite{Asa04,Dat04,Ume05,Iid06,Jak07,Aar07}.
Potential models provide a simple and intuitive framework 
for the study of quarkonium properties at finite temperature 
but the main problem is the determination of the effective potential.
At zero temperature, the use of effective potentials
and their connection to the underlying field theory is well
established but at finite temperature the issue is still open
\cite{Sim05,Lai07a,Lai07b}.
Different potential models based upon the lattice free and internal energies
have a fair agreement with the results from the lattice 
studies \cite{Won07,Alb07} whereas the calculations of 
Euclidean correlators using a variety of potential models were
not able to reproduce the temperature dependence of the lattice correlators
\cite{Moc06,Cab06,Alb07}. It was not until recently that effective 
field theory (EFT) approaches have been developed to justify
the use of potential models at finite temperature
because the EFT techniques are suited for handling 
systems with well separated  energy scales. 
Two types of scales appear {\em viz.} at T=0:
the mass of the heavy quark, $m_Q$, the typical momentum 
transfer ($m_Q v_Q$, and the binding energy ($m_Q v_Q^2$), 
and at $T \ne 0$: the temperature $T$, 
the Debye mass $m_D (=g T)$, and $g^2 T$~\cite{Brambilla:2008cx,
Brambilla:2010vq}. Since the relative velocity ($v_Q$) 
is very small ($v_Q \sim \alpha_s<1$)~\cite{Brambilla:2004jw, 
Pineda:2011dg}, both types of scales are well separated. 

Nowadays it is understood that the dissociation of heavy quarkonia is 
not due to the Debye screening of the potential, but is rather overtaken
by the thermal width obtained from the imaginary part
of the potential~\cite{Laine:2006ns,Brambilla:2008cx,
Laine:2008cf}.  
There are two different processes in QCD, which contribute to the 
thermal width: the first process is the inelastic parton scattering
mediated by the space-like gluons, 
when the Debye mass is greater than the binding energy. This process
is often referred as Landau damping and is the principal source
of dissociation mechanism at weak coupling.
The  second process is the gluon-dissociation process which 
corresponds to the decay of a color singlet state into a 
color octet induced by a thermal gluon and is dominant 
when the binding energy is greater than the Debye mass~\cite{
Brambilla:2013dpa}. 
Recently one of us derived both the real and the imaginary component 
of the potential 
by correcting both the perturbative and nonperturbative (string) 
terms of the potential at T=0 in real-time formalism. We found that after inclusion
of confining string term,
the (magnitude) imaginary component becomes larger and hence
contribute more to the thermal width, compared to the 
medium-modification of the perturbative term 
alone~\cite{Vineet:PRC2009,Lata:PRD2013,Lata:PRD2014}
However, the above discussions are limited to the weak coupling techniques.

The large and anisotropic collective flow and its
strong interaction with the hard probes at collider experiments indicate that
the matter created must be described by QCD in a regime of strong interactions. 
In the strong coupling regime, the effect of an imaginary part 
in usual potential models has been addressed in \cite{Petreczky:2010tk}, 
and also in the T-matrix approach through the Schwinger-Dyson equations
\cite{Mannarelli:2005pz}. The imaginary part has also been calculated on the 
lattice \cite{Rothkopf:2011db,Burnier:2012az}, where
the spectral functions have been extracted, using the limited sets of data.
This fact, together with the intrinsic technical difficulties of 
lattice calculations, somehow limits the reliability
of the results obtained so far, and also their scope. 
Therefore some complementary methods for strong coupling at finite 
temperature are desirable. The AdS/CFT conjecture provides such an 
alternative~\cite{Maldacena:1998adth,Witten:1998,Gubser:1998}
for a class of non-abelian thermal gauge field 
theories, which gives an analytic access to the strong coupling 
regime of finite 
temperature gauge field theories in the limit of large number of 
colors. This is achieved by mapping the problems at strong coupling
onto calculable problems in the supergravity limit of a
dual string theory. The background metric will be described by
a curved five-dimensional AdS spacetime containing a black hole
with its horizon lying in the fifth dimension.
Although the AdS/CFT correspondence is not directly applicable 
to QCD, but the universality of the result
for the shear viscosity-to-entropy ratio
with the prediction of hydrodynamics in HIC~\cite{Teaney:2003kp} 
motivates to calculate 
the heavy quark potential at finite temperature.

The expectation values of Wilson loops contain
gauge invariant information about
the nonperturbative physics of non-abelian gauge field theories.
In fact, the expectation value of a particular time-like Wilson loop 
serves to define the potential between a static quark and antiquark 
at finite temperature. The first calculation was done by 
Maldacena~\cite{Maldacena:1998im}
for $ {\cal N} = 4 $ SYM for T=0 and was later extended to finite temperature 
in~\cite{Brandhuber:1998bs,Rey:1998bq}.
The potential is obtained by extremizing the world sheet of open
string attached to the heavy quark and its anti quark located at the 
boundary of ${\rm{AdS}}_5$ space in the background of ${\rm{AdS}}_5$ space
\cite{Maldacena:1998im} and in the background of ${\rm{AdS}}_5$ 
black-hole metric~\cite{Brandhuber:1998bs,Rey:1998bq},
for determining the potentials at $T=0$ and $T\ne 0$, respectively.
The problem with pure AdS background is that it does not capture the 
renormalisation group running, like in QCD and does not give
any confining term in the potential.
In this work we use the Ouyang-Klebanov-Strassler-Black-Hole (OKS-BH) geometry
\cite{KS} explored recently in \cite{Mia:NPB2010,Mia:PRD2010,Mia:PLB2011}, 
which captures the logarithmic running of 
the coupling constants at far IR and is almost conformal at 
UV and calculate both real and imaginary parts of the potential 
at finite temperature by analytically continuing the string
configurations into the complex plane. Our work is organised as follows.
In Section 2, we have revisited the OKS-BH
geometry in brief and in Section 3, we employ this geometry to
construct the Nambu-Goto action for obtaining the potential.
Finally we conclude in Section 4.

\section{Geometry}
Let us suppose a gauge theory living in $3 + 1$ dimensions 
defined at a particular energy scale $\Lambda $. It can be 
thought of as being defined on a $3 + 1$ dimensional slice in 
a higher dimensional space and if we stack up all such slices 
together we can 
have a continuous family of gauge theories labelled by 
$\Lambda$ . We can imagine this continuous family as a single 
theory in five dimensions with $\Lambda$ becoming the fifth 
dimension. The flow along this energy scale 
is known as renormalisation group (RG) flow. Now if the theory 
is conformal i.e. does not flow with energy ($\Lambda$), it can 
be described on the boundary of a five-dimensional space. 
In other words a conformal field theory (CFT) can be 
described conveniently on the boundary of pure Anti-de Sitter (AdS) space 
\cite{Maldacena:1998adth}. But if the theory has a RG flow we 
cannot describe the
full theory simply by describing it on the boundary of some 
five dimensional space and we need to look differently at 
different energy scales. In particular for large N QCD, we have 
conformal behaviour at UV and logarithmic running of coupling 
at IR and we need to look differently at these two regimes. 
The geometries that can account for such a RG flow have been 
constructed and discussed in detail in 
\cite{Mia:PRD2010,Mia:NPB2010}. We shall briefly review them here. 
\paragraph{ }
It has been seen that if we embed D branes in a geometry, the 
gauge theory living on these D branes may show logarithmic RG 
flow. 
A particular model that does this is the Klebanov-Strassler (KS) warped
 conifold construction \cite{KS}. The gravity dual of 
this construction, which is a warped deformed conifold with three 
form type IIB fluxes, captures the RG flow of the gauge theory. The 
corresponding gauge theory is confining in the far IR but is 
not 
asymptotically free. Additionally, in KS picture the quarks are 
all in the bi-fundamental representations of the two possible UV gauge 
groups and they eventually cascade away in the far IR. Also this 
construction is for zero temperature gauge theories.
What we need is a dual gravity theory that allows quarks in the 
fundamental representation at high temperature.
\paragraph{}
The problem of having a fundamental quark can be resolved by 
inserting $N_f$ D7 branes in the KS geometry. This is a subtle issue 
and so far we only know how to insert coincident D7 branes in the 
Klebanov-Tseytlin background \cite{KS}. The resulting background 
is the Ouyang  background \cite{KS} that has
all the type IIB fluxes switched on, including the axio-dilaton.
Now to 
switch on a non zero temperature we need to insert a black hole into 
this background and the Hawking temperature of the black hole will 
correspond to the gauge theory temperature.
Combining all physics ingredients together 
the metric in OKS-BH geometry looks like 
\begin{equation}
ds^2 = {1\over \sqrt{h}}
\Big[-g_1(u)dt^2+dx^2+dy^2+dz^2\Big] +\sqrt{h}\Big[g_2^{-1}(u) du^2+ d{\cal M}_5^2\Big]
\end{equation}
where $h$ is the warp factor, $g_i(u)$ are the black-hole 
factors, $u$ denotes the extra dimension and 
$d{\cal M}_5^2$ is the metric of warped resolved-deformed 
conifold. 
 
The above  picture works well at IR. Now we need to see that the theory 
should become conformal and free at UV. In other words we want  
AdS-Schwarchild geometry in asymptotic limit. It is clear that  
we cannot use the pure OKS-BH background and we need to introduce
the appropriate UV cap to it. As a consequence, 
the above metric will receive corrections due to the UV cap as
\begin{equation} \label{eq:correction}
ds^2 = {1\over \sqrt{h}}
\Big[-g(u)dt^2+dx^2+dy^2+dz^2\Big] +\sqrt{h}\Big[g(u)^{-1}g_{uu}du^2+ g_{mn}dx^m dx^n\Big]
\end{equation}
where we have set black hole factors $ g_1(u) = g_2(u) = g(u)$. The corrections $ g_{uu} $ 
are of the form $ u ^{-n} $ and appear in the metric because the 
existence of axio-dilaton and the seven-brane sources tell us that the unwarped metric 
may not remain Ricci flat. These may be written as  
\begin{equation}
g_{uu} = 1 + \sum_{i = 0}^{\infty} \frac{a_{uu,i}}{u^i}
\end{equation}
where $ a_{uu,i}$ are the coefficients similar to $a_i$ and can be solved for exactly as 
shown in \cite{Mia:PRD2010}.
\\
The
warp factor, $h$ can be obtained as
\begin{equation*}
 h ~= ~\frac{L^4}{u^4}\left[1+\sum_{i=1}^\infty\frac{a_i}{u^i}\right]~~~ 
\end{equation*}
where $a_i $ are coefficients of ${\cal O}(g_sN_f)$ and can be solved 
for exactly as shown in \cite{Mia:PRD2010} and $L$
 denotes the curvature of space. 
This metric reduces to OKS-BH in IR and is asymptotically $AdS_5$ x 
$M_5$ in the UV. It describes the geometry all the way from the IR to the
UV.
\\
With the change of coordinates $ z = 1/u $, we can rewrite the metric as
\begin{equation}\label{eq: metric}
ds^2 ~=~ g_{\mu\nu} dX^\mu dX^\nu ~=~
 A_n z^{n-2}\left[-g(z)dt^2+d\overrightarrow{x}^2\right]
~ +~\frac{
B_l z^{l}}{
 A_m z^{m+2}g(z)}dz^2+\frac{1}
{A_n z^{n}}~ds^2_{{M}_5},
\end{equation}
where $ds^2_{{\cal M}_5}$ is the metric of the internal space 
and $A_n$'s are the coefficients that can be extracted from the 
$a_i$'s as follows: 
\begin{equation}
{1\over \sqrt{h}}~ = ~ {1\over L^2 z^2 \sqrt{a_i z^i}} \equiv  
A_n z^{n-2} ~=~ {1\over L^2 z^2}\left[a_0  
-{a_1z\over 2} + \left({3a_1^2\over 8a_0} - {a_2\over 2}\right)z^2 
+ \cdot \cdot \cdot \right] \quad,
\end{equation}
which gives $ A _0 = {a_0\over L^2},  A_1 = -{a_1\over 2L^2},  
A_2 = {1\over L^2}\left({3a_1^2\over 8a_0} - {a_2\over 
2}\right)$ and so on. Note that since $a_i$'s for  $i \ge 1$ 
are of ${\cal O}(g_sN_f)$ and $L^2 \propto \sqrt{g_sN}$, so 
in the limit $ g_s N_f 
\rightarrow 0$ and $ N \rightarrow \infty $ all $ 
A_i$'s for $ i \ge 1$ are very small. The $u^{-n}$ corrections mentioned 
above in Eq.(\ref{eq:correction}) are accommodated via $ B_l z^l$ series which is given by  
\begin{equation}
B_l z^l = 1 + a_{zz,i} z^i
\end{equation}

In fact the complete picture can be divided into three regions. Region 1 is 
the IR region where we have pure OKS-BH geometry. Region 3 is the UV 
region where UV cap has been added. And region 2 is the interpolating 
region between UV and IR. 
The background for all these three regions and the process of adding 
UV cap has been described in full details in \cite{Mia:PRD2010}. Also the RG 
flow associated with these regions and the corresponding field theory 
realizations have been discussed in \cite{Chen:PRD2013}. We shall not go into 
the complete details here and will use the metric given
in Eq.(\ref{eq: metric}) in extremizing the action 
to calculate the potential in the next section.

\section{Heavy quark potential from Gravity}
One of the most important gauge invariant quantities defined 
in non-Abelian $SU(N_c)$ gauge theories 
\cite{Wilson:1974sk,wilsonloop} 
is the Wilson loop
\begin{equation}
\label{eq:wilsonloop}
W({\cal{C}}) = \frac{1}{N_c}\textrm{tr} \, P \, \exp{ \left[ i g\oint_C A_{\mu}
dx^{\mu} \right]},
\end{equation}
where $\cal{C}$ is a closed loop in a 4-dimensional spacetime, 
which is usually taken as a rectangular loop of spatial
and temporal extensions  $r$ and $\mathcal{T}$, respectively.
$P$ indicates path-ordering, $g$ is the coupling, $A_{\mu}$ is 
the non-Abelian gauge field potential operator. The trace (tr) is 
performed over the fundamental representation of the gauge field. 
The vacuum expectation value ('o') of the Wilson
loop gives the desired heavy quark potential $V_{Q \bar Q}$:
\begin{equation}
\label{eq:wilsonrec}
\lim_{\mathcal{T} \to \infty}\langle W(C) \rangle_0 \sim e^{i \mathcal{T} V_{Q\bar{Q}}(r)},
\end{equation}
At finite temperature, the thermal average of the Polyakov loop
correlator ($\langle {\rm tr}\,
\bold{L}^{\dagger}(r)\,{\rm tr}\,\bold{L}(0)\rangle$) gives the free energy 
of a heavy $Q\bar{Q}$ pair \cite{McLerran:1980pk} at finite temperature,
which is quite often taken as the heavy quark potential at finite temperature
on the lattice. However, the potential at finite temperature can also
be obtained from the Wilson loop's expectation value 
evaluated in a thermal state of the gauge theory :
\begin{equation}
\label{eq:wilsonrec}
\lim_{\mathcal{T} \to \infty}\langle W(C) \rangle 
\sim e^{i \mathcal{T} V_{Q\bar{Q}}(r,T)}
\end{equation}
Maldacena was the first to calculate the rectangular Wilson loop in 
the vacuum of strongly coupled $\mathcal{N}=4$ SYM 
theory~\cite{Maldacena:1998im} and was later 
calculated in finite temperature from the
correlator of two Polyakov loops in \cite{Brandhuber:1998bs,
Rey:1998bq}. 
According to the gauge/gravity prescription \cite{Maldacena:1998im}, the 
expectation value of $W(C)$ in a strongly coupled gauge theory 
dual to a theory of gravity is
\begin{equation}
\label{eq:wilsongaugegravity}
\langle W(C) \rangle \sim Z_{str},
\end{equation}
where $Z_{str}$ is the generating functional of the string in the 
bulk, having the loop $C$ at the boundary. In the classical gravity 
approximation
\begin{equation}
\label{eq:classicalpartition}
Z_{str} \sim e^{i S_{str}},
\end{equation}
where $S_{str}$ is the classical string action propagating 
in the bulk evaluated at an extremum. 

In general this heavy quark potential can have an imaginary 
part~\cite{Lai07a,Beraudo:NPA2008,Petreczky:2010tk,Brambilla:2010vq,
otherrefsImV, Rothkopf:2011db}, which contributes to the thermal decay width 
and is related to the imaginary part of the gluon 
self-energy. There are two processes responsible for the
thermal width, namely the Landau damping and the 
singlet to color octet transition. 
In AdS/CFT framework, people have obtained the non-zero imaginary part 
for large separation (beyond a critical separation, $r_c$) by 
analytically continuing the string configurations into the 
complex plane~\cite{Yavier:PRD2008}. Recently the imaginary contribution 
has been obtained from the thermal fluctuation around the bottom of the sagging 
string in the bulk that connects the heavy quarks located at the 
boundary~\cite{Noronha:2009da,Noronha:2013JHEP}, where the thermal fluctuation 
causes the maximum length of the sagging classical string to cross the
horizon of the black hole and makes the potential absorptive.
In both cases, the potential is obtained by extremizing the world sheet of open
string attached to the heavy quark pair located at the 
boundary of ${\rm{AdS}}_5$ space. Recently a real-time complex 
potential $V_{Q\bar{Q}}$(t,r) is derived 
from the Wilson loop in the background of AdS black hole
by the analytic continuation from the imaginary time to 
the real time~\cite{Hatsuda:2013adscft}, where
the complex string configuration 
has been interpreted as a real string which is moving rather 
than being static.

Following the background geometry described 
above in section 2, the Nambu-Goto action can be defined as
\begin{equation}
S_{NG} = \frac{1}{2 \pi} \int d \sigma d \tau \sqrt{-
det \left[(g_{\mu \nu} + \partial_\mu \phi \partial_\nu 
\phi)\partial_a X^{\mu} \partial_b X^{\nu}\right]}
\end{equation}
where $\phi$ is background dilaton field. It is responsible for breaking 
of conformal symmetry of theory and is given by 
\begin{equation}
\phi = \log g_s - g_s D_{n + m_o} z^{n + m_o}
\end{equation}
Using the parametrization $ X^{\mu} = (t,x,0,0,z),~\tau = 
t,~\sigma = x $ and $\partial_a = \frac{\partial}{\partial 
\tau},\partial_b = \frac{\partial}{\partial \sigma} $,
we extremize the open string 
worldsheet attached to a static quark at $x = + r/2 $ and an 
anti-quark at $x = -r/2$. Using the metric given above in 
Eq. (\ref{eq: metric}), the Nambu-Goto action can be rewritten
as 
\begin{eqnarray}
S_{NG} &=& \frac{\mathcal{T}}{2 \pi} \int_{-\frac{r}{2}}^{\frac{r}{2}} 
\frac{dx}{z^2} \nonumber\\
&&\sqrt{(A_n z^n)^2 g(z) +\left( B_l z^l + 2 
g(z)g_s^2(n + m_o)D_{n + m_o}(l + m_o)D_{l + m_o} A_k z^{k + 
n+ l + 2m_o}\right) (z')^2}~~~~~
\end{eqnarray}
where $ g(z)= 1-\frac{z^4}{z_h^4}$. $z_h = \frac{1}{\pi T }$ is the black hole 
horizon.  By defining
\begin{equation}
B_mz^m=B_l z^l + 2 g(z)g_s^2(n + m_o)D_{n + m_o}(l + 
m_o)D_{l + m_o} A_k z^{k + n+ l + 2m_o}~,
\end{equation} 
the above action can be written in a closed form
\begin{equation}
S_{NG} = \frac{\mathcal{T}}{2 \pi} \int_{-\frac{r}{2}}^{\frac{r}{2}} dx 
\sqrt{(A_n z^n)^2 \left(\frac{1}{z^4}-\frac{1}{z_h^4} \right) 
+ B_m z^m \frac{(z')^2}{z^4} }. 
\end{equation}
The Nambu-Goto action can also written as an integral over $z$,
\begin{equation} \label{eq: action}
S_{NG} = \frac{\mathcal{T}}{\pi} \int_0^{z_{max}} \frac{d z}{z'} 
\sqrt{(A_n z^n)^2 \left(\frac{1}{z^4}-\frac{1}{z_h^4} \right) 
+ B_m z^m \frac{(z')^2}{z^4} } 
\end{equation}
Since this action does not depend explicitly on $x$, so the corresponding 
Hamiltonian will be a constant of motion, i.e., 
\begin{equation}
H = z' \frac{\partial L}{\partial z'} - L = C_0~({\rm{say}})
\end{equation} 
Thus the constant $C_0$ can be determined as 
\begin{equation*}
\frac{1}{L}\left( (z')^2 \frac{B_m z^m}{z^4}-L^2\right) = - \left(\frac{1}
{z^4} - \frac{1}{z_h^4}\right) \frac{(A_n z^n)^2}{L} = C_0 
\end{equation*}
Using the fact that at $ z = z_{max} $, $z^\prime = 0$ ,
where $z_{max}$ is the maximum of the string coordinate along 
fifth dimension, we can find out $z^\prime$ by simple algebra
\begin{equation} \label{eq: z der}
z' = \frac{d z}{d x} = \frac{\sqrt{z_h^4 - z^4}(A_n z^n)}{z_h^2 
\sqrt{B_m z^m}} \left( \frac{(z_h^4 -z^4)(A_n z^n)^2 
z_{max}^4}{(z_h^4 -z_{max}^4)(A_n z_{max}^n)^2 z^4} - 1 
\right) ^\frac{1}{2} 
\end{equation}
Integrating both sides, the above equation yields the separation
$r$ as 
\begin{eqnarray}
r = \frac{2 z_h^2 \sqrt{z_h^4 -z_{max}^4}(A_n z^n_{max})}
{z_{max}^2} && \int_0^{z_{max}} dz \frac{z^2 \sqrt{B_m z^m}}{{(z_h^4 
- z^4)} (A_n z^n)^2}  \times \nonumber\\
&& \left(1 - \frac{(z_h^4 -z_{max}^4)(A_n z_{max}^n)^2 z^4}
{(z_h^4 -z^4)(A_n z^n)^2 z_{max}^4} \right)^{-\frac{1}{2}} 
\end{eqnarray}
Expanding the square root and keeping only the first term  because 
the coefficients $A_i$'s are small, the separation ($r$) becomes
\begin{equation} \label{eq: r-z}
r = \frac{2 z_h^2 \sqrt{z_h^4 -z_{max}^4}(A_n z^n_{max})}
{z_{max}^2} I,
\end{equation}
where the integral $I$ is defined by
\begin{equation}
I = \int_0^{z_{max}}dz \frac{z^2 \sqrt{B_m z^m}}{{(z_h^4 - 
z^4)} (A_n z^n)^2}
\end{equation}

\par
Now we obtain the action by substituting $z^\prime$ from Eq.(\ref{eq: z der}) into
Eq.(\ref{eq: action})
\begin{equation*}
S_{NG} = \frac{\mathcal{T}}{\pi} \int_0^{z_{max}} \frac{d z \sqrt{B_m 
z^m}}{z^2}\left(1 - \frac{(z_h^4 -z_{max}^4)(A_n z_{max}^n)^2 
z^4}{(z_h^4 -z^4)(A_n z^n)^2 z_{max}^4} \right)^{-\frac{1}{2}} 
\end{equation*}
Expanding the square root and keeping only the first two terms (since the 
coefficients $A_i$'s are small), the action is obtained as
\begin{equation}
S_{NG} = \frac{\mathcal{T}}{\pi} \left[ \int_0^{z_{max}} \frac{d z 
\sqrt{B_m z^m}}{z^2} ~ + ~ \frac{1}{2}\frac{(z_h^4 -z_{max}^4)(A_n 
z_{max}^n)^2}{z^4_{max}} I \right]
\end{equation}
After substituting the integral $I$ in terms of $r$, the action becomes
\begin{eqnarray}
S_{NG}  &=&  \frac{\mathcal{T}}{\pi} \int_0^{z_{max}} \frac{d z \sqrt{B_m 
z^m}}{z^2} ~+~ \frac{\mathcal{T}}{4 \pi} \frac{\sqrt{z_h^4 -z_{max}^4}}
{z_h^2 z^2_{max}} A_n z^n_{max} r\nonumber\\
&\equiv & S^{1} + S^{2} 
\end{eqnarray}

The first term in the action, $S^1$ diverges in the lower limit of the
integration, so we regularize it by integrating from $\epsilon$, instead
of 0 to $z_{max}$~\cite{Mia:PRD2010} and identifying the divergent 
term in the integral
\begin{equation*}
S^{1} = \frac{\mathcal{T}}{\pi} \int_{\epsilon}^{z_{max}} 
\frac{d z \sqrt{B_m z^m}}{z^2} \quad ,
\end{equation*}
where we take $A_0 = 1$ and $A_1 = 0$ and similarly 
$ B_0 = 1$ and $B_1 = 0 $. Let us also assume for simplicity that 
the higher coefficients $A_i $ and $B_i $ are very small for $i \geq 
3$ and can be neglected. This simplification reduces the series $ A_n z^n$ to 
$ 1 + A_2 z^2 $ and 
$ B_m z^m $ to $ 1 + B_2 z^2 $. This will also simplify all 
the expressions and help us to keep an analytic control on the 
equations. In the limit of small coefficients, $S^1$ becomes
\begin{equation}
S^{1} = \frac{\mathcal{T}}{\pi} \int_{\epsilon}^{z_{max}} \frac{d z (1 + \frac{B_2}{2} z^2)}{z^2}
\end{equation}
After integrating and separating into finite and divergent terms, $S^1$ 
becomes in the limit $\epsilon \rightarrow 0$
\begin{equation*}
S^{1} = \lim_{\epsilon \rightarrow 0} 
\frac{\mathcal{T}}{\pi}\left(-\frac{1}{z_{max}} + \frac{1}
{\epsilon} + \frac{B_2}{2}(z_{max}-\epsilon) \right)
\end{equation*}
Therefore subtracting the divergent piece in the limit
$\epsilon \rightarrow 0$ from the 
action, we obtain the renormalised action 
\begin{equation}
S^{ren}_{NG} = \frac{\mathcal{T}}{4 \pi} \frac{\sqrt{z_h^4 -z^4_{max}}}
{z_h^2 z^2_{max}} (1 ~+~ A_2 z^2_{max}) r ~+~ \frac{\mathcal{T}}{\pi}\left(-\frac{1}{z_{max}} ~+~ \frac{B_2}{2} z_{max} \right)
\end{equation}
and hence the potential is given by
\begin{eqnarray}
V_{Q\bar{Q}} &=& \lim_{\mathcal{T} \rightarrow \infty} \frac{S^{ren}_{NG}}{\mathcal{T}} \nonumber\\
&=&  \frac{1}{4 \pi} \frac{\sqrt{z_h^4 -z^4_{max}}}
{z_h^2 z^2_{max}} (1 + A_2 z^2_{max}) r + 
\frac{1}{\pi}\left(-\frac{1}{z_{max}} +\frac{B_2}{2} z_{max} \right) \quad,
\end{eqnarray}
which is a function of both $z_{\rm{max}}$ and r. We will therefore
express $z_{\rm{max}}$ as a function of r from Eq.(\ref{eq: r-z}) and then
plug in to the above potential.
To do that we first concentrate on the integral $I$, which is 
simplified into, after neglecting the terms beyond second order:
\begin{equation} \label{eq: I}
I = \int_0^{z_{max}}dz \frac{z^2 \sqrt{1 + B_2 z^2}}{{(z_h^4 - 
z^4)} (1 + A_2 z^2)^2}
\end{equation} 
In the limit of small coefficients,
this integral turns out to be
\begin{eqnarray}
I &=& -\frac{1}{8 (-1 + A_2^2 z_h^4)^2} \left[ \frac{2(2 A_2 - B_2)
(-1 + A_2^2 z_h^4) z_{max}}{(1 + A_2 z^2_{max})} \right. \nonumber\\
&-& \left. \frac{2 (6 A_2
- B_2 + A_2^2 z_h^4 ( 2 A_2 - 3 B_2)}{\sqrt{A_2}} \tan ^{-1}
(\sqrt{A_2} z_{max}) \right. \nonumber\\
&-&\left. \frac{2 (1 + A_2 z_h^2)^2 (-2 + B_2
z_h^2)}{z_h} \tan^{-1} \frac{z_{max}}{z_h} \right. \nonumber\\
&+& \left. \frac{(-1 + A_2
z_h^2)^2)(2 + B_2 z_h^2)}{z_h} \log \left( \frac{z_h -
z_{max}}{z_h + z_{max}} \right) \right]
\end{eqnarray}
Next we will find the solution for $ r$ in two limits, namely
$z_{max} >> z_h $ and $z_{max} << z_h $. For $z_{max} >> z_h $, 
the separation $r$ from Eq.(\ref{eq: r-z}) becomes
\begin{equation*}
r = 2 i z_h^2\left(1 -\frac{z_h^4}{2 z_{max}^4}\right) (1 + 
A_2 z^2_{max}) I (z_{\rm{max}} >> z_h)
\end{equation*}
Using the asymptotic expansions of the functions appeared in the integral $I$:
\begin{equation*}
\tan^{-1} z (z >> 1) = \frac{\pi}{2} - \frac{1}{z} + \frac{1}
{3 z^2} - \frac{1}{5 z^5} + O[\frac{1}{z}]^7
\end{equation*} 
\begin{equation*}
\log \left( \frac{(\frac{z_h}{z_{max}}-1)}{(\frac{z_h}
{z_{max}} + 1)}\right) = i \pi - 2 \frac{z_h}{z_{max}} - 
\frac{2}{3} \frac{z_h^3}{z_{max}^3} - \frac{2}{5}\frac{z_h^5}
{z_{max}^5} - O[\frac{z_h}{z_{max}}]^7~~,
\end{equation*}
the separation $r$ thus becomes as a function of $z_{\rm{max}}$
\begin{eqnarray}
\label{rlargezmax}
\frac{r}{2 i z_h^2} &=& -(c_1 + i c_2) A_2  z^2_{max} - (c_1 + i c_2) 
+ \frac{B_2}{6 A_2 z_{max}} +  \frac{1}{2}\frac{z_h^4 A_2}{z^2_{max}}
 (c_1 + i c_2) \nonumber\\
&+&  \frac{(6 
A_2 - B_2)}{30 A_2^2 z^3_{max}} +\frac{1}{2}\frac{z_h^4 }
{z_{max}^4} (c_1 + i c_2) 
\end{eqnarray}
where $c_1$ and $c_2$ are defined as 
\begin{eqnarray} \label{eq: c}
c_1 &=&  \frac{\pi}{8} \left[\frac{1}{z_h}\left(\frac{2-
B_2 z_h^2}{(A_2 z_h^2 - 1)^2} \right) - \frac{6 A_2 - B_2 + A_2^2 z_h^4(2 A_2 
-3 B_2)}{\sqrt{A_2} (-1 + A_2^2 z_h^4)^2 } \right] \\ \nonumber 
 c_2 &=&  \frac{\pi}{8 z_h} \frac{2 + B_2 z_h^2}{(A_2 
z_h^2 + 1)^2}
\end{eqnarray}
We now invert the series (\ref{rlargezmax}) to obtain $ z_{max} $ in terms of 
$r$ as, 
\begin{eqnarray} \label{eq: zmax-r1}
z_{max} &=& \frac{1}{\sqrt{2 A_2 z_h^2(-i c_1 + c_2)}} \sqrt{r} 
~-~ \sqrt{ \frac{(-i c_1 + c_2) z_h^2}{2 A_2}} \frac{1}{\sqrt{r}} 
~-~ \frac{i}{6}\frac{B_2}{A_2} \frac{z_h^2}{r} \nonumber\\ 
&+&\frac{((-i c_1 + c_2) z_h^2)^{3/2} (-1 + 2 A_2^2 z_h^4)}{2 
\sqrt{2 A_2}} \frac{1}{r^{3/2}} \nonumber\\
&+& \frac{2 (3 A_2 + 2 B_2) (-c_1 - i c_2)}{15 A_2} \frac{z_h^4}{r^2} 
+ O[\frac{1} {r}]^{5/2} ,
\end{eqnarray}
which shows that the string coordinates become complex
and hence the potential becomes imaginary. We have 
checked numerically that this situation corresponds to 
$ r > r_c$, where $r_c$ is some critical separation and for this 
geometry it turns out to be $0.53 z_h $. However, the authors of 
\cite{Rey:1998bq,Brandhuber:1998bs} abandoned the solution for 
$r > r_c$ but the authors in \cite{Yavier:PRD2008} suggested 
that the complex-valued saddle points beyond $r_c$ (=$0.87 z_h$) 
may be interpreted as the quasi-classical configurations in the classically 
forbidden region of string coordinates, analogous to the 
method of complex trajectories 
used in quasi-classical approximations to quantum mechanics \cite{LL3}.
The complexification of the string coordinates simply indicates that
the saddle point of the integral over string coordinates becomes
complex. According to the standard AdS/CFT prescription
\cite{Maldacena:1998im}, in the large-$N_c$ large-$\lambda$ limit the
integral over string coordinates is still dominated by the saddle
point, even if it is complex so the saddle point approximations
is still valid~\cite{Yavier:PRD2008}. Recently authors 
in \cite{Hatsuda:2013adscft} estimated the critical
length scale $0.62 z_h$, in the same sense of \cite{Yavier:PRD2008},
with an interpretation of the complex string configuration 
as a real string which is moving rather than being static.

In this limit ($z_{\rm{max}} >> z_h$), the potential reduces to
\begin{equation}
V_{Q \bar{Q}} = \frac{1}{\pi} \left( \frac{B_2}{2} z_{max} - 
\frac{1}{z_{max}} \right) - \frac{i r}{4 \pi z_h^2} \left(A_2 
z^2_{max} + 1 - \frac{A_2}{2} \frac{z_h^4}{z^2_{max}} - 
\frac{1}{2} \frac{z_h^4}{z_{max}^4} \right)   
\end{equation}
Here we could have chosen either $ +i $ or $-i $ in the second 
term above. To select the correct sign we note that time 
evolution operator in quantum mechanics is $ e^{-iEt} \sim 
e^{ Im [E]t } $. To maintain unitarity we demand that the 
Im[V(r)] should be negative. When we demand this we find that 
we should use a negative sign here.  

Finally we substitute for $z_{max}$ in terms of r from 
Eq.(\ref{eq: zmax-r1}), to obtain the potential as a function of $r$
and $T$ only: 
\begin{equation}
\begin{split}
V_{Q \bar{Q}} (r,T) = - \frac{i}{8 \pi (-i c_1 + c_2)} \frac{r^2}
{z_h^4} ~+~ \frac{5}{12 \sqrt{2}} \frac{B_2}{\sqrt{A_2 (-i c_1 ~+~  c_2)}} \frac{\sqrt{r}}{ \pi z_h}   \\ - \frac{9}{20 \sqrt{2}} 
\frac{(4 A_2 + B_2) \sqrt{-i c_1 + c_2}}{\sqrt{A_2}}  
\frac{z_h}{\pi \sqrt{r}}  ~-~ \frac{13 i}{72} \frac{B_2 ^2}{A_2 
\pi }\frac{z_h^2}{r}  
\end{split}
\end{equation}
We can split it into real and imaginary parts, where the
real part is given by (a dimensionless variable, $\hat{r} = r T $)
\begin{eqnarray}
Re[V_{Q \bar{Q}}] (\hat{r},T) &\stackrel{z_{\rm{max}} \gg z_h}{\simeq}&
\frac{1}{8} \frac{c_1}{c_1^2 + c_2^2} 
\pi^3 T^2 \hat{r}^2 + \frac{5}{12 \sqrt{2}} \frac{B_2}
{\sqrt{A_2}} \frac{1}{\sqrt{c_1^2 + c_2^2}} 
\sqrt{\frac{\sqrt{c_1^2 + c_2^2} + c_2}{2}} \sqrt{T \hat{r}} \nonumber\\
&-&\frac{9}{20 \sqrt{2}} \frac{(4 A_2 + B_2) }{\sqrt{A_2}} 
\sqrt{\frac{\sqrt{c_1^2 + c_2^2} + c_2}{2}} \frac{1}{\pi^2 
\sqrt{T \hat{r}}}  
\end{eqnarray}
For small $r$, the first two terms (positive) tend to  zero while 
the third term becomes (negative) large, hence the potential
is highly attractive.  For large r, the first two terms (positive) become 
larger while the third  term (negative) becomes smaller, hence the
potential goes to continuum.
As the temperature increases the first two terms becomes
larger, hence the potential becomes quickly zero for smaller r.
This can be understood
by the fact that as the temperature increases, screening becomes 
stronger.

The imaginary part upto leading order in $r$ is  given by
\begin{eqnarray}
\label{imaginary}
Im[V_{Q \bar{Q}}] (\hat{r},T) &=& -\frac{1}{8} \frac{c_2}{c_1^2 + c_2^2} \pi^3 
T^2 \hat{r}^2 \nonumber\\
&+& \frac{5}{12 \sqrt{2}} \frac{B_2}
{\sqrt{A_2}} \frac{1}{\sqrt{c_1^2 + c_2^2}} 
\sqrt{\frac{\sqrt{c_1^2 + c_2^2} - c_2}{2}} \sqrt{T \hat{r}},
\end{eqnarray}
which shows that the imaginary part vanishes 
for small r and agrees with the perturbative result.

We will now find the potential for the 
other extreme limit i.e.., very small $ z_{max} $ 
and $z_{max} << z_h $. In this limit, the separation $r$ becomes
\begin{equation} \label{eq: r-z1}
r = \left(\frac{2 z_h^4}{z_{max}^2} ~-~ z_{max}^2 \right) (1 + 
A_2 z_{max}^2) I (z_{max} << z_h)
\end{equation} 
Using the expansion of $\tan^{-1}$ and $\log$ functions in the integral 
$I(z_{max} << z_h)$
\begin{equation*}
\tan^{-1} z (z << 1) = z - \frac{1}
{3} z^3 + \frac{1}{5} z^5 + O[z]^7
\end{equation*} 
\begin{equation*}
\log \left( \frac{1-\frac{z_{max}}{z_h}}{1+ \frac{z_{max}}
{z_h }}\right) = - 2 \frac{z_{max}}{z_h} - 
\frac{2}{3} \frac{z_{max}^3}{z_h^3} - \frac{2}{5}\frac{z_{max}^5}{z_h^5} 
+ O[\frac{z_{max}}{z_h}]^7~~,
\end{equation*}
the separation $r$ can be expressed in a series of $z_{\rm{max}}$ from Eq. (\ref{eq: r-z1})
\begin{equation}
r = \frac{2}{3} z_{max} + \frac{3 B_2 - 2 A_2}{15} z_{max}^3 + \frac{-5 + 6 A_2^2 z_h^4 - 9 A_2 B_2 z_h^4}{105 z_h^4} z_{max}^5,
\end{equation}
which will eventually be inverted to express $z_{max}$ in terms of 
$r$ as
\begin{equation} \label{eq: zmax-r}
\begin{split}
z_{max} = \frac{3 r}{2} ~+~ \frac{27}{80} ( 2 A_2 - 3 B_2) r^3 ~+~ O[r]^5 
\end{split}
\end{equation}

In this limit, the potential in Eq.(25) reduces to 
\begin{equation}
V_{Q \bar{Q}} = \frac{1}{\pi} \left( \frac{B_2}{2} z_{max} - 
\frac{1}{z_{max}} \right) + \frac{ r }{4 \pi } \left( \frac{1}{z_{max}^2} + A_2 - \frac{1}
{2} \frac{z_{max}^2}{z_h^4} - \frac{A_2}{2} \frac{z_{max}^4}{z_h^4}  \right)   
\end{equation}
After plugging $z_{max}$ from (\ref{eq: zmax-r}), the potential 
in $z_{max} <<z_h$ limit becomes 
\begin{equation}
\begin{split}
Re[V_{Q \bar{Q}}] (r,T) \stackrel{z_{\rm{max}} \ll z_h}{\simeq}
-\frac{5}{9 \pi r} 
+ \frac{9 (A_2 + B_2)}{20 \pi} r + \frac{27}{224 \pi} 
\left( \frac{(2 A_2 - B_2)(2 A_2 + 7 B_2)}{50} - 
\frac{1}{z_h^4} \right) r^3 + O[r]^5
\end{split}
\end{equation}
The above form of potential looks like a (Cornell) 
potential at $T=0$,
apart from a tiny $r^3$ temperature-dependent term, which can be 
understood by the fact that  in small distance limit ($rT<<1$), 
$Q \bar {Q}$ pair does not see the medium.
Note that there is no imaginary part in this limit, as expected. Since 
we have the potential in both asymptotic limits, we 
can interpolate them to find a general expression for the
potential
\begin{equation}
\label{realinter}
Re[V_{Q \bar{Q}}] (r,T) \approx \frac{5}{9 \pi r} \frac{(r - r_0)^3}{r_0^3}
\end{equation}
where $ r_0 $ is the new scale and is given by 
\begin{equation} \label{eq: rnot}
r_0 = \left( \frac{40 (c_1^2 + c_2^2)}{9 c_1 \pi^4 T ^4} \right) ^{1/3} 
\end{equation}
This formula reduces to -$1/r$ form  for $ r << r_0 $ 
and reduces to $r^2$ form for $ r >> r_0 $. The parameter $r_0$ can 
be interpreted as the screening length and it's dependence on the 
temperature can be seen from Eq. (\ref{eq: c}) and (\ref{eq: rnot}). 
We have evaluated both real (\ref{realinter}) and imaginary parts 
(\ref{imaginary}) of the potential
for increasing temperatures (Fig.1), to see how the potential
obtained via gauge-gravity duality in the strong coupling regime 
gets screened with the increase in temperature.
\section{Conclusion and Discussions} 
We have obtained  a heavy quark potential at finite temperature 
in a dual gravity which is somewhat closer to QCD than pure 
AdS geometry in the sense that it accounts for RG flow. For the small separation  limit 
$(rT << 1)$ (which implies either $r \rightarrow 0$ or $T 
\rightarrow 0$), the potential looks like a Cornell potential at $T = 0$, with a small 
higher-order confining term, whereas at large separation $(rT >> 1)$, the potential gets 
screened, having a quadratic term, a $\sqrt{r}$  term and an attractive $1/ \sqrt{r}$ 
term.
The quadratic term  is probably originated from the non-dominant 
cubic term at small separation. In addition to it, the potential 
develops an imaginary term above some separation, $ r_c = 0.53 z_h $.
The difference in the estimate of critical separations in  other
calculations \cite{Yavier:PRD2008, Hatsuda:2013adscft}
may be simply due to the different world sheet configuration taken.

Moreover we found that the screening of potential is in 
the form of a power law (which means $Re[V(r)]$ falls off as some 
power of $1/rT$) in our case, unlike the exponential falloff due to  
debye screening in the long distance in which $Re[V(r)] \sim \exp(-m_D r) $.  

\section{Acknowledgements}
We would like to thank Yuri V. Kovchegov for his extremely helpful 
comments and suggestions and also for patiently explaining his 
paper. We would also like to thank Keshav Dasgupta for looking 
through the calculations. B. K. P is thankful to the CSIR project (No. 03/1215), 
Government of India, for the 
financial assistance..

\begin{figure}
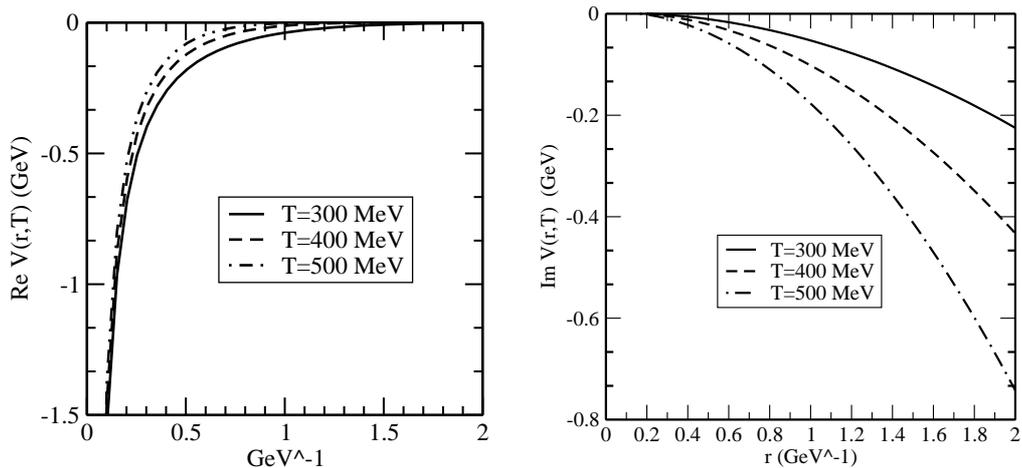

\begin{center}$
\begin{array}{cc}
\includegraphics[keepaspectratio,width=2.5in,height=2.5in]{real_inter.eps} 
\hspace{0.25in}
\includegraphics[keepaspectratio,width=2.5in,height=2.5in]{binoy_ima.eps} \\
\end{array}$
\end{center}
\caption{\footnotesize Variations of the real and imaginary part of 
the potential with the separation $r$, where the value of both 
coefficients $A_2$ and $B_2$ are taken as 0.24.}
\end{figure}

\end{document}